\documentclass[eqsecnum,aps,twocolumn,showpacs,superscriptaddress]{revtex4}

\usepackage{graphicx}
\usepackage{times}

\begin{document}

\title{Giant Spin-splitting in the Bi/Ag(111) Surface Alloy}
\author{Christian R. Ast}
\affiliation{Ecole Polytechnique F\'ed\'erale de Lausanne (EPFL),
Institut de Physique des Nanostructures, CH-1015 Lausanne,
Switzerland}\affiliation{Max-Planck-Institut f\"ur
Festk\"orperforschung, 70569 Stuttgart, Germany}
\author{Daniela Pacil\'e}
\affiliation{Ecole Polytechnique F\'ed\'erale de Lausanne (EPFL),
Institut de Physique des Nanostructures, CH-1015 Lausanne,
Switzerland}
\author{Mihaela Falub}
\affiliation{Ecole Polytechnique F\'ed\'erale de Lausanne (EPFL),
Institut de Physique des Nanostructures, CH-1015 Lausanne,
Switzerland}
\author{Luca Moreschini}
\affiliation{Ecole Polytechnique F\'ed\'erale de Lausanne (EPFL),
Institut de Physique des Nanostructures, CH-1015 Lausanne,
Switzerland}
\author{Marco Papagno}
\affiliation{Ecole Polytechnique F\'ed\'erale de Lausanne (EPFL),
Institut de Physique des Nanostructures, CH-1015 Lausanne,
Switzerland}
\author{Gero Wittich}
\affiliation{Max-Planck-Institut f\"ur Festk\"orperforschung,
70569 Stuttgart, Germany}
\author{Peter Wahl}
\affiliation{Max-Planck-Institut f\"ur Festk\"orperforschung,
70569 Stuttgart, Germany}
\author{Ralf Vogelgesang}
\affiliation{Max-Planck-Institut f\"ur Festk\"orperforschung,
70569 Stuttgart, Germany}
\author{Marco Grioni}
\affiliation{Ecole Polytechnique F\'ed\'erale de Lausanne (EPFL),
Institut de Physique des Nanostructures, CH-1015 Lausanne,
Switzerland}
\author{Klaus Kern}
\affiliation{Ecole Polytechnique F\'ed\'erale de Lausanne (EPFL),
Institut de Physique des Nanostructures, CH-1015 Lausanne,
Switzerland} \affiliation{Max-Planck-Institut f\"ur
Festk\"orperforschung, 70569 Stuttgart, Germany}

\begin{abstract}
Surface alloying is shown to produce electronic states with a very
large spin-splitting. We discuss the long range ordered
bismuth/silver(111) surface alloy where an energy bands separation
of up to one eV is achieved. Such strong spin-splitting enables
angular resolved photoemission spectroscopy to directly observe
the region close to the band edge, where the density of states
shows quasi-one dimensional behavior. The associated singularity
in the local density of states has been measured by low
temperature scanning tunneling spectroscopy. The implications of
this new class of materials for potential spintronics applications
as well as fundamental issues are discussed.
\end{abstract} 

\maketitle

Manipulating the electron spin without employing magnetic fields
is a vision that lies at the heart of spintronics. The spin-orbit
(SO) interaction --- which couples orbital and spin degrees of
freedom --- provides the basis for spin manipulation by means of
electric fields. It plays a vital role in various device proposals
in spin-based quantum information technology
\cite{Wolf01,Datta90,Koga,Ohe}. One proposal for a spin transistor
\cite{Datta90}, for example, relies on the spin-precession of a
propagating electron due to a SO-induced spin splitting
\cite{Rashba,Henk03}. However, external electric fields are
typically not strong enough to induce an appreciable phase shift
within the electron's mean free path. Internal electric fields,
e.\ g.\ induced by spacial inversion asymmetry, are much stronger,
yet the spin-splitting in semiconductors, which are the materials
of choice in spintronics, is smaller than what is found for
metallic surface states \cite{LaShell96}.

Clean surfaces of elemental metals show a trend of strong atomic
spin-orbit coupling leading to a large spin-splitting of their
surface states, which can be further enhanced by the adsorption of
adatoms \cite{Rotenberg,Hochstrasser02,Krupin}. This is a
promising path to create a new class of nanoscale structures where
morphology and chemistry are used to tune the spin-splitting of
interface states. Surface alloying in particular provides
interesting opportunities as the adatoms \textit{replace}
substrate atoms of the topmost monolayer in their lattice sites
creating a new two-dimensional electronic structure.

In this report we demonstrate that the band structure of the
bismuth/silver surface alloy grown on a Ag(111) substrate exhibits
a giant spin splitting. Angular-resolved photoemission
spectroscopy (ARPES) has been employed to map the
characteristically offset parabolic bands. This large splitting
offers experimental access to a region near the band maximum where
the density of states diverges and the spin orientation changes
its sense of rotation. Low temperature scanning tunneling
microscopy/spectroscopy (STM/STS) has been employed to study the
topography and the local density of states (LDOS) of the alloy. A
distinct peak in the measured LDOS shows, as predicted by theory,
the quasi one-dimensional van Hove singularity at the band edge.
This introduces an experimental approach for identifying spin
splitting by STS.

\begin{figure}
\includegraphics[width=3.4in]{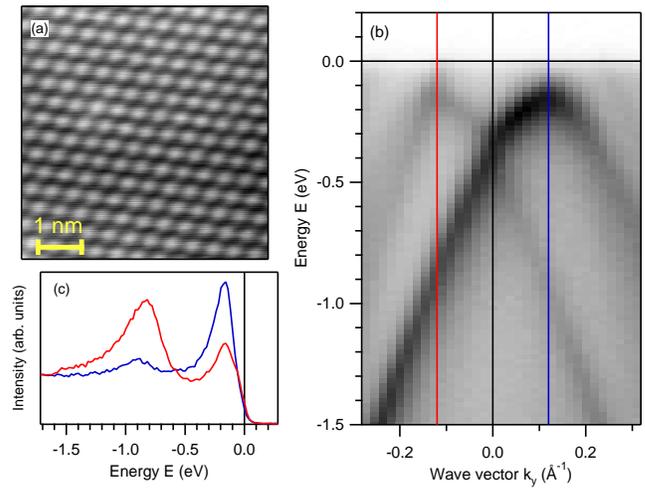}
\caption{(color online) (a) Topography by STM of the long-range
ordered Bi/Ag(111) surface alloy (Bias voltage: -3\,mV; Tunneling
current: 1\,nA). The alloy is grown in ultra-high vacuum by
depositing one third of a monolayer of bismuth onto a clean
Ag(111) surface at a temperature of 400\,K to yield a
$\sqrt{3}\!\times\!\sqrt{3}R30^{\circ}$ reconstruction. (b) ARPES
band structure image near the $\overline{\Gamma}$-point, i.\ e.\
the center of the surface Brillouin zone, showing the spin-split
bands of the Bi/Ag(111) surface alloy. The intensity scale is
linear with light and dark areas corresponding to low and high
intensity, respectively. (c) Energy distribution curves to the
left and right of the $\overline{\Gamma}$-point as indicated by
the respective lines in (b).} \label{Start}
\end{figure}

The ARPES measurements were done at room temperature and LN2
(77\,K) temperatures using 21.2\,eV photons (HeI) in ultra-high
vacuum ($2\cdot 10^{-10}\,$mbar). Sample preparation was done {\it
in situ}. The energy and angular resolution of the analyzer were
better than 10\,meV and $\pm 0.015\,\mbox{\AA}^{-1}$. The light is
partially polarized with the polarization vector within a mirror
symmetry plane perpendicular the (111) surface of the crystal.
Photoelectrons were collected within the mirror plane for the
geometry of Fig.\ \ref{Start}. The STM/STS measurements were done
at 6\,K in ultra-high vacuum ($1\cdot 10^{-10}\,$mbar) with {\it
in situ} sample transfer and preparation.

Figure \ref{Start}(a) shows a topographic scan of the long-range
ordered hexagonal surface alloy (5\,\AA\ lattice constant) taken
by STM. Bright spots correspond to Bi atoms, each of which is
surrounded by six Ag atoms, as indicated by the model in the
inset. Structural details will be published elsewhere. The
corresponding band structure measured by ARPES near the
$\overline{\Gamma}$-point at the center of the surface Brillouin
zone (SBZ) (Fig.\ \ref{Start}(b)) shows two identical and nearly
parabolic bands with negative effective mass. They replace the
nearly free electron-like surface state of the bare Ag(111)
surface and accommodate the $p$-electrons donated by the Bi atoms.
Remarkably, their maxima are shifted to the left and right of
$\overline{\Gamma}$. The offset $k_0$ increases from
$0.12$\,\AA$^{-1}$ at room temperature to $0.16$\,\AA$^{-1}$ at
liquid nitrogen temperatures (77\,K), which is about 22\,\% of the
SBZ.

The band on the right exhibits higher intensity. To better
visualize this effect Fig.\ \ref{Start}(c) shows two energy
distribution curves to the left and right of $\overline{\Gamma}$.
This asymmetry is the result of the parity of the electronic wave
functions combined with the (linear) polarization of the exciting
beam \cite{Henk03}. The observation of two symmetrically offset
bands with an asymmetric distribution of intensities is, as for
the surface state of the clean Au(111) surface \cite{Henk}, a
clear indication of spin-orbit induced spin-splitting.

These observations can be qualitatively understood on the basis of
a simple nearly free electron model. The Hamiltonian describing
the spin-orbit coupling at the surface is
\cite{Henk03,Petersen00}:
\[H_{\textrm{SOC}}=\alpha (\vec{e}_z\times \vec{k})\cdot\vec{\sigma}\]
where $\alpha$ is proportional to an effective electric field. The
vector $\vec{e}_z$ is perpendicular to the surface and
$\vec{k}=(\vec{k}_{||}, k_{\perp})$ is the electron wave vector,
with components parallel and perpendicular to the surface. The
energy dispersion is:
\[E(\vec{k}_{||})=\frac{\hbar^2}{2m^*}(k_{||}-k_0)^2 + E_0\]
where $m^*$ is the effective mass, $k_0$ is the offset by which
the parabola is shifted away from $\overline{\Gamma}$ and a
function of $\alpha$. $E_0$ is an offset in energy. The energy
dispersion is rotationally symmetric; it only depends on the
magnitude of $\vec{k}_{||}$.

\begin{figure}
\includegraphics[width=3.4in]{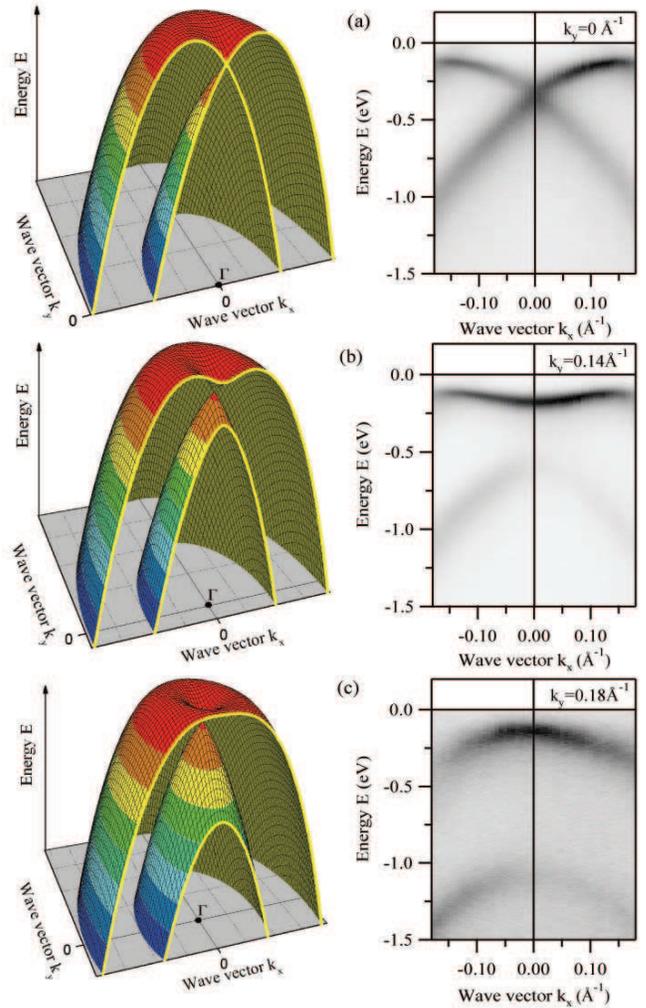}
\caption{(color online) (a)-(c) Calculated energy dispersion in
the nearly free electron model (left) and experimental band
structure images at different positions in k-space (right). The
cuts are perpendicular to the image in Fig.\ \ref{Start}(c) which
leads, for our experimental geometry, to a homogeneous intensity
distribution over the two branches. The yellow lines in the
calculation correspond to the measured section.} \label{Cuts}
\end{figure}

The energy dispersion near $\overline{\Gamma}$ is visualized
(energy vs. wave vector $k_x$ and $k_y$) in the left panels of
Fig.\ \ref{Cuts}(a)-(c). The right panels show a set of
experimental band structure images. Fig.\ \ref{Cuts}(a) shows a
cut through the center of the Brillouin zone. The bands cross at
$\overline{\Gamma}$ and reach their maxima at $\pm k_0$ near the
edges of the measured images. Away from $\overline{\Gamma}$ along
the $k_y$-axis (Fig.\ \ref{Cuts}(b)), the bands no longer cross.
The upper band is rather flat while the lower band disperses with
a parabola-like shape. In Fig.\ \ref{Cuts}(c), at
$k_y=0.18\,\mbox{\AA}^{-1}$, beyond the band maximum, both bands
show a parabola-like dispersion with a separation in energy of
935\,meV!

\begin{figure}
\includegraphics[width=3.4in]{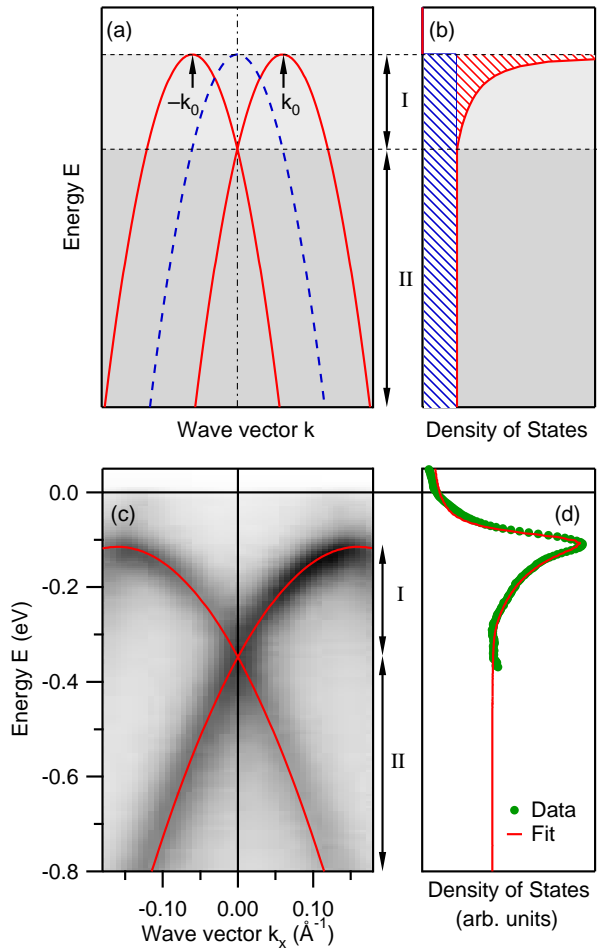}
\caption{(color online) Direct comparison of theory and
experiment: (a) Calculated energy dispersion in the nearly free
electron model with (red lines) and without (blue dashed line)
spin-orbit coupling. (b) The corresponding density of states. (c)
ARPES map of the band dispersion (red line) as a guide to the eye.
(d) Local density of states measured by STS (green dots) with a
fit from the nearly free electron model (red line).}
\label{Theory}
\end{figure}

Two qualitatively different energy regions can be identified which
are detailed in Fig.\ \ref{Theory}(a). Region I reaches from the
band maximum to the crossing of the two inner branches; region II
reaches from this crossing point to lower energies. The main
difference between these region concerns the density of states
$D(E)$, which is easily evaluated analytically:
\[D(E)=\int \delta(E-E(\vec{k_{||}}))\,\frac{d\vec{k}_{||}}{4\pi^2} = \]
\[= \left\{ \begin{array}{ll}
\displaystyle\frac{|m^*|}{\pi\hbar^2} \frac{k_0}{\sqrt{2m^*(E-E_0)/\hbar^2}} & ;\ E \in \textrm{Region I} \\
\displaystyle\frac{|m^*|}{\pi\hbar^2}  & ;\ E \in \textrm{Region II}\\
\displaystyle 0 & ;\ \textrm{elsewhere}\\
\end{array}\right. \]
The density of states (DOS) in Fig.\ \ref{Theory}(b) is constant
in region II like in the two-dimensional free electron model
without spin-orbit splitting. In region I it follows a
$1/\sqrt{E}$-behavior reminiscent of the van Hove singularity in
one-dimensional models. At the band maximum the DOS diverges and
then drops to zero. This is a signature of the spontaneous
symmetry breaking which occurs for any finite $k_0$ turning the
two-dimensional point-like band maximum for $k_0=0$ into a
quasi-one-dimensional ring-like maximum.

Going from theory to experiment, the image in Fig.\
\ref{Theory}(c) shows the band structure measured by ARPES. The
two red lines are the two shifted parabolic bands in the model
with an effective mass $m^*=-0.4m_e$ as a guide to the eye. The
extent of Region I is about 220\,meV. Fig.\ \ref{Theory}(d)
displays a dI/dV spectrum measured by STS (green dots) which is
proportional to the LDOS of the sample. It shows a distinct peak
which can be identified with the singularity in the calculated DOS
in Fig.\ \ref{Theory}(b). The red line in Fig.\ \ref{Theory}(d) is
a fit to the data by convoluting the DOS with a Lorentzian
(40\,meV full width at half maximum) to account for lifetime
effects and experimental broadening. The fitted spin splitting
parameter is $k_0=0.13\,\mbox{\AA}^{-1}$ in good agreement with
ARPES.

The singularity at the band edge is a distinct feature of a
spin-split band in a two-dimensional electron gas (2DEG).
Therefore, the peak in the LDOS introduces a useful method for
identifying the spin-splitting of energy bands by STS. It could
not be observed in, for example, the Au(111) surface state because
there the splitting is not as pronounced. If the width of Region
I, determined by $k_0$, is too narrow, broadening effects
immediately dampen the singularity which will only appear as a
step in the LDOS \cite{Kliewer}. However, the asymmetric
singularity may cause a small shift of the leading edge as well as
a steeper slope. On the other hand, standing waves in dI/dV maps
are often inconclusive in determining the spin splitting of a 2DEG
because electrons with opposite spin do not interfere
\cite{Petersen00}.

\begin{figure}
\includegraphics[width=3.4in]{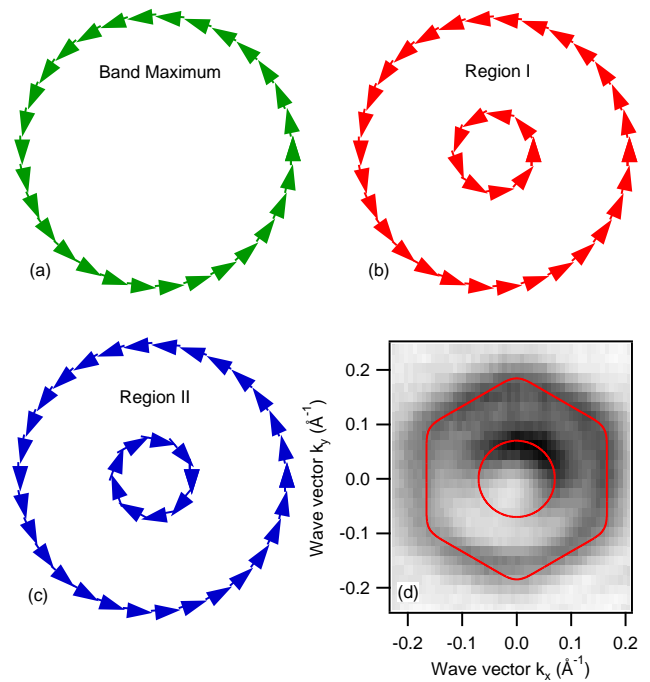}
\caption{(color online) Directions of spin rotation of the bands
according to the nearly free electron model at the band maximum
(a), in Region I (b), and Region II (c). (d) Constant energy slice
of the spin split bands in Region I at an energy of $-180$\,meV
measured by ARPES. The red lines indicate the contours of the
energy bands as a guide to the eye.} \label{Circles}
\end{figure}

The spin orientation calculated from the spin-orbit Hamiltonian
$H_{\textrm{SOC}}$ is displayed schematically in Fig.\
\ref{Circles}(a)-(c) for three different energies, at the band
maximum, within Region I, and within Region II. The spins lie
within the plane of the surface where the sense of rotation
depends on the direction of the electric field. At the band
maximum only one constant energy contour is visible (we assume a
counterclockwise spin orientation). Within Region I two contours
of the same spin orientation can be associated to the inner and
the outer branches of the energy bands. Crossing over to Region II
the inner contour changes its sense of spin rotation. This
scenario is different from what has been previously observed in
spin-split bands where only Region II is observable experimentally
\cite{LaShell96,Henk}. The large spin-splitting in the
bismuth/silver surface alloy provides access to Region I with a
qualitatively different spin configuration. The constant energy
image in Fig.\ \ref{Circles}(d) measured by ARPES shows two
concentric contours at an energy of $-180$\,meV, which is well
within Region I. The inner contour is nearly circular, while the
outer one is hexagonal due to its stronger interaction with the
lattice potential. Qualitatively, however, the picture described
in Figs.\ \ref{Circles}(a)-(c) concerning the spin-rotation holds
true for the bismuth/silver surface alloy.

It is interesting to note that in the present system the high-Z
element is not the substrate but the doping material. Moreover,
neither clean Ag(111) ($k_0=0.004\,\mbox{\AA}^{-1}$)
\cite{Popovic} nor the pristine Bi(111) surfaces
($k_0\approx0.05\,\mbox{\AA}^{-1}$) \cite{Koroteev} exhibit such a
strong spin splitting effect. We conclude that the formation of a
surface alloy where adatoms are integrated in the topmost surface
layer can lead to a strong enhancement of the effect. Model
calculations give insight into this problem \cite{Petersen00}, but
clearly first principle calculations are required to better
understand the spin-dependent band structure in such interfaces as
a function of chemical and structural parameters.

The giant spin-splitting observed in the Bi/Ag(111) surface alloy
is not a unique phenomenon particular to this combination of
materials but rather a property of a new class of materials. In
particular, experiments on the Pb/Ag(111) surface alloy have shown
an almost equally large spin-splitting with the band maximum in
the unoccupied states. This suggests that tuning of the Fermi
level across the spin-split bands could be achieved by doping the
bismuth/silver alloy with lead atoms. Preliminary work, which we
have done, supports this idea. The ability of tuning the Fermi
level through the different regions of the spin-split bands will
offer an ideal playground to test fundamental ideas.

On a broader perspective, surface alloys could be tailored to
specific applications. Ordered thin films of silver on a silicon
surface have already been grown successfully \cite{Katayana}.
Extending this to the present study, spin-split bands in a surface
alloy on a semiconductor substrate becomes a real possibility with
exciting perspectives in spintronics. Applying the spin-splitting
parameters of the bismuth/silver alloy to the problem of the spin
transistor, we find for the spin precession that a phase
difference of $\Delta\theta=\pi$ is reached after a distance of
$L=\Delta\theta/k_0=2.6\,$nm \cite{Datta90}, which is about two
orders of magnitude smaller than for a semicondutor.

We gratefully acknowledge discussions with D.\ Malterre and H.\
Brune. C.R.A. acknowledges funding from the Emmy-Noether-Program
of the Deutsche Forschungsgemeinschaft. The work at the EPFL was
supported by the Swiss National Science Foundation through the
MaNEP NCCR.

\end{document}